\documentclass{aa}  
\usepackage{graphicx}
\usepackage{txfonts}
\usepackage{longtable}
\usepackage{natbib}
\bibpunct{(}{)}{;}{a}{}{,}
\usepackage{url}
\newcommand{\fermi}{\textit{Fermi}}

\begin{document} 

\title{VLBA observations of radio faint \fermi -LAT sources above 10\,GeV.}

   \author{R.~Lico\inst{1}\fnmsep\thanks{Email: rocco.lico@unibo.it}, M.~Giroletti\inst{1}, 
   M.~Orienti\inst{1} \and F.~D'Ammando\inst{1,2}.
           }

   \institute{INAF Istituto di Radioastronomia, via Gobetti 101, 40129 Bologna, Italy.
\and Dipartimento di Fisica e Astronomia, Universit\`a di Bologna, via Ranzani 1, 40127 Bologna, Italy.
              }

\date{Received ; accepted }

  \abstract
   {The first \fermi-LAT High-energy source catalog (1FHL), containing $\gamma$-ray sources detected above 10\,GeV, is an ideal sample to characterize the physical properties of the most extreme $\gamma$-ray sources. 
   }
   {We investigate the pc scale properties of a sub-sample of radio faint 1FHL sources with the aim to confirm the proposed blazar associations, by revealing a compact high brightness temperature radio core, and we propose new low-frequency counterparts for the unassociated $\gamma$-ray sources (UGS). Moreover, we increase the number of 1FHL sources with high resolution observations to explore the possible connection between radio and $\gamma$ rays at E >10\,GeV. 
   }
   {We observed 84 1FHL sources, mostly blazars of High Synchrotron Peaked (HSP) type, in the northern sky with the Very Long Baseline Array (VLBA) at 5\,GHz. These sources lack high resolution radio observations and have at least one NRAO VLA sky survey counterpart within the 95\% confidence radius. 
   For those sources without a well identified radio counterpart we exploit the VLBA multiple phase-center correlation capability to discern among the possible low-frequency candidates.
   }
   {For $\sim$93\% of the sources of our sample we reveal a compact high brightness temperature radio core, supporting their proposed blazar association. The vast majority of the detected sources are radio weak, with a median VLBI flux density value of 16.3 mJy. 
   For the detected sources we obtain an average brightness temperature of the order of 2 $\times 10^{10}$ K.
   We find a compact component for 16 UGS, for which we propose a new low-frequency association.
   }
   { We find brightness temperature values which do not require high Doppler factors, and are in agreement with the expected values for the equipartition of energy between particles and magnetic field. We find strong indications about the blazar nature of all of the detected UGS, for which we propose new low-frequency associations. 
The characterization of the physical properties of this emerging population is relevant in view of the construction of the new generation Cherenkov Telescope Array.
   }

  \authorrunning{R.\ Lico et al.}
\titlerunning{VLBA observations of radio faint \fermi -LAT sources above 10\,GeV.}
   \maketitle 
 

\section{Introduction}

The vast majority of high energy (HE, 100\,MeV $<E<$ 100\,GeV) and very high energy (VHE, $E>0.1$\,TeV) $\gamma$-ray sources are associated with radio loud objects, typically blazars, i.e.\ flat spectrum radio quasars (FSRQs) or BL Lac type objects (BL Lacs) \citep[e.g. ][]{Acero2015, Ackermann2015}. Depending on the position of the peak of the synchrotron component of their Spectral Energy Distribution (SED), blazars are further classified as Low-, Intermediate-, or High-Synchrotron Peaked (LSP, ISP, HSP, respectively) sources, characterized by a synchrotron peak frequency $\nu_\mathrm{peak}$ (Hz) such that $\log \nu_\mathrm{peak}<14$, $14<\log \nu_\mathrm{peak}<15$, and $\log \nu_\mathrm{peak}>15$, respectively \citep{Abdo2010a}.

In the MeV/GeV domain, the Large Area Telescope (LAT) on board the \textit{Fermi Gamma-ray telescope} (\fermi) has provided a deep, uniform sky survey detecting as many as 3,033 sources in the \fermi\ third source catalog \citep[3FGL, ][]{Acero2015}. The results from \fermi\ show that both types of blazars are common HE emitters, with FSRQs having softer spectra and being generally more luminous than BL Lacs. Moreover, \fermi\ clearly revealed the existence of a highly significant correlation between the radio flux density and the $\gamma$-ray energy flux \citep{Ackermann2011}. However, observations from Imaging Atmospheric Cherenkov Telescopes (IACTs) detect preferentially BL Lac sources, in particular of the HSP type, and no evidence of a correlation between radio and VHE emission has been reported so far. 
The different demographics of the VHE population is explained by two main factors: FSRQs have softer spectra and are more distant, which decreases their VHE emission due to extragalactic background light (EBL) absorption. 
The lack of a VHE-radio correlation may be due to the IACTs operational mode. IACTs operate in pointing mode with a limited sky coverage, and they preferentially observe sources in a peculiar state. All of these limitations introduce a strong bias in VHE catalogs, and it is difficult to assess any possible radio-VHE correlation.
Nonetheless, it is likely that some physical effect may also be at work.

The First \fermi-LAT Catalog of Sources above 10\,GeV \citep[1FHL,][]{Ackermann2013} is an ideal resource for addressing the connection between HE and VHE emission. The 1FHL is based on LAT data accumulated during the first 3 years of the \fermi\ mission, providing a deep and uniform all-sky survey. It contains 514 sources, 
of which $\sim$76\% are statistically associated with Active Galactic Nuclei (AGN), $\sim$11\% are sources of Galactic nature (pulsars, supernova remnants, and pulsar wind nebulae), while $\sim$13\% remain unassociated.
Various observational campaigns were dedicated to search for the low-frequency counterpart of the unassociated $\gamma$-ray sources (UGS) detected by \fermi\ \citep{Nori2014, Massaro2013b, Giroletti2016}. 
Recently, \citet{Schinzel2015} found 76 new low-frequency associations of $\gamma$-ray sources between 5 and 9\,GHz, by detecting parsec-scale emission through Very Long Baseline Interferometric (VLBI) observations.

We are now investigating 1FHL sources at $\delta>0^\circ$ with the use of high angular resolution VLBI radio observations. The goals of this project are multi-fold: support the proposed blazar associations by revealing a compact high brightness temperature radio core; search for new counterparts for UGS; increase the size of the population of $E>10$\,GeV sources with high angular resolution observations; explore the existence of a correlation between VLBI and $E>10$\,GeV emission on a sample as large and unbiased as possible.

In this paper we present the first results of this project. 
From the whole 1FHL catalog we extracted a sample of 269 sources in the northern sky with a radio counterpart in the NRAO VLA Sky Survey \citep[NVSS, ][]{Condon1998} within the 95\% confidence radius (r95), by using TOPCAT software \citep{Taylor2005}. We call this the 1FHL-n sample. For 185 out of these 269 sources Very Long Baseline Array (VLBA) archival observations are already available. Eighty-four sources ($\sim$31\%) have never been observed with the VLBA, and 21 of them are UGS. 
We focus our attention on this these 84 1FHL sources (see Table~\ref{tab_sources_log}).
For the sources with an association we aim to confirm their low-frequency association and to study their parsec scale properties. For the UGS we want to investigate their nature, possibly detecting a compact radio source associated with them.

The paper is organized as follows: in Sect.~\ref{sec.observations}, we present the observations and data reduction procedures; we show the results in Sect.~\ref{sec.results}; we discuss and summarize the results in Sect.~\ref{sec.discussion} and in Sect.~\ref{sec.conclusions}, respectively. 
In a following paper, we will present a detailed statistical analysis of the entire 1FHL sample, based on new and archival data.

Throughout the paper, we use a $\Lambda$CDM cosmology with $h = 0.71$, $\Omega_m = 0.27$, and $\Omega_\Lambda=0.73$ \citep{Komatsu2011}.
The radio spectral index is defined such that $S(\nu) \propto \nu^{-\alpha}$ and the $\gamma$-ray  photon index $\Gamma$ such that  $dN_{\rm photon}/dE \propto E^{-\Gamma}$.

\addtocounter{table}{1}

\begin{table}
\centering
\caption{Observations details.}
\label{obs_log}
\begin{tabular}{lll}
\hline
\hline
Observing date & Experiment Code & Stations log \\
\hline
2013 Dec 9  & S6340A & No FD \\
2013 Oct 3  & S6340B & No LA \\
2013 Oct 4  & S6340C & No LA \\
2013 Sep 30 & S6340D & No LA  \\
2013 Nov 22 & S6340E & No FD \\
2013 Dec 7  & S6340F & No FD \\
2015 Jul 10 & S6340G & No PT \\
2015 Dec 8  & S6340H & - \\
\hline
\end{tabular} 
\tablefoot{Station codes: FD -- Fort Davis, LA -- Los Alamos, PT -- Pie Town.}
\end{table}

\addtocounter{table}{1}

\begin{figure*}
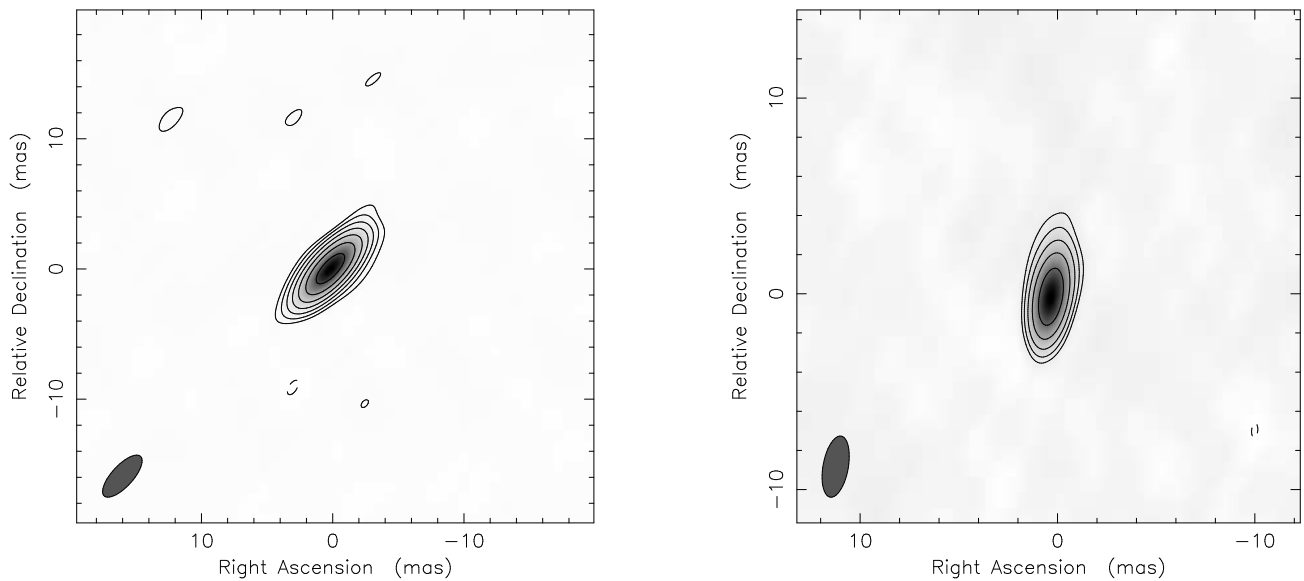

\includegraphics[bb=70 154 470 641, width=0.85\columnwidth, angle=-90, clip]{J0043.ps}
\includegraphics[bb=70 154 470 641, width=0.85\columnwidth, angle=-90 ,clip]{J0648.ps} \\
\caption{5\,GHz VLBA image of the radio counterparts of the $\gamma$-ray sources 1FHL J0043.7+3425 (left panel), classified as a FSRQ, and 1FHL J0648.9+1516 (right panel), classified as a BL Lac. The beam size is 1.6 mas $\times$  4.1 mas and 1.3 mas $\times$ 3.2 mas, respectively. Levels are drawn at $(-1, 1, 2, 4...) \times$ the lowest contour (that is, at 1.2 mJy/beam for the source 1FHL J0043.7+3425 and at 1.4 mJy/beam for the source 1FHL J0648.9+1516) in steps of 2. The noise level is 0.31 mJy and 0.37 mJy, respectively.} 
\label{maps}
\end{figure*}

\section{Observations and data reduction} 
\label{sec.observations}
We performed the VLBA observations between 2013 September 30 and December 7, with two additional observing epochs between 2015 July 10 and December 8. We divided the sample into eight observing blocks (Table~\ref{obs_log}). The observations were carried out at a central frequency of 5\,GHz in full polarization, with a total bandwidth of 256\,MHz divided into eight 32 MHz sub-bands in each polarization, with a recording rate of 2 Gbps. Observations were carried out in phase reference mode, with a duty cycle of 5 minutes (4 min on target and 1 min on calibrator), repeated six times, resulting in a net observing time of $\sim24$ minutes per source. The total observing time for this project is 48 hours. 
As reported in Table~\ref{obs_log}, during some observing epoch some antennas did not work properly because of technical problems. 

Even if the angular resolution at $E>10$\,GeV ($<0.15^{\circ}$) is smaller than at lower $\gamma$-ray energies ($<3.5^{\circ}$ at $E>100$\,MeV), it still is much larger than the typical field of view of VLBI observations (milliarcsecond scale). Moreover, there are 18 sources for which there is not a well identified radio counterpart, but rather a set of a few possible candidates (between 2 and 5). For these sources we exploit the capability of the DiFX correlator \citep{Deller2011} to correlate the data at multiple phase centers, one for each candidate radio counterpart. Thanks to this feature it is possible to observe all of the radio sources within the r95 region with one single pointing. However, some of them are located at a distance from the pointing center that is comparable to the 25 m single dish primary beam at 5\,GHz ($\sim 8$ arcmin).

We carried out a full calibration by using the software package Astronomical Image Processing System \citep[AIPS, ][]{Greisen2003} with the following steps: phase calibration by applying ionospheric and Earth orientation parameters correction; amplitude calibration based on gain curves and measured system temperatures; parallactic angle correction; removal of residual instrumental phase and delay offset by using the pulse-cal tones table; global fringe fitting on all of the calibrator sources, and transfer of the solutions to the targets. 

After this, we split phase-referenced calibrated visibility data, averaging data in frequency within each sub-band but not across IFs, and without any time average (the correlator integration time was 4 second). This provided a nominal field of view of 6.2 arcseconds.
With these single source data sets, we started the editing and imaging process in DIFMAP \citep{Shepherd1997}. Since the phase tracking centers were based on NVSS observations, which for faint sources can be as inaccurate as a few arcseconds, we often had to image wide sky regions before finding a significant source in the image plane. We determined the accurate positions and shift from the phase tracking center for all of the detected sources (see Table~\ref{tab_sources_log}). We then went back to the multi file in AIPS and for each source we corrected the position by using the task CLCOR.

For the six sources (1FHL J0338.4+1304, 1FHL J0516.4+7351, 1FHL J1244.9+5708, 1FHL J1315.0+2346, 1FHL J1942.8+1034 and 1FHL J2223.4+0104) whose calibrators have some extended structure, after the above mentioned calibration procedure, we imaged the calibrator in DIFMAP, and we produced a model that we use in AIPS for a couple of self-calibration cycles with the task CALIB. 
The source 1FHL J0930.4+8611 is quite strong (S>200 mJy/beam) and we performed a global fringe-fit on the source itself.

After splitting the single source visibility datasets, we started a final cycle of imaging and self calibration in AIPS, obtaining for each source the peak flux density and  the deconvolved angular size by using the task JMFIT. Finally, for the sources with a significant offset from the pointing center, we corrected for primary beam attenuation by using the widefield VLBI calculator\footnote{\url{http://www.atnf.csiro.au/people/Emil.Lenc/Calculators/wfcalc.php}}.

For 11 sources (marked with an asterisk in Table~\ref{tab_sources_log}) we observed the NVSS source which is closer to the 1FHL source centroid, but which is not coincident with the proposed low-frequency association in the 1FHL. For this reason we exclude these 11 sources from our analysis. We also exclude the supernova remnant source 1FHL J1911.0+0905 and the final sample discussed in this paper consists of 72 sources.

\section{Results} 
\label{sec.results}

\subsection{VLBI properties and detection rate}
\label{sec_detection_rate}
We detect radio emission on VLBI scales for 67/72 sources, which corresponds to an overall detection rate of 93\%. For blazars the detection rate is 100\% (51/51), while for the UGS it is 76\% (16/21).
The significance level for the detection of these sources is always $\gg5\sigma$. 
The VLBI coordinates, the 10-500\,GeV energy flux ($S_{\gamma}$), the 1.4\,GHz NVSS flux density ($S_{1.4, NVSS}$), the 5\,GHz VLBA flux density ($S_{5, VLBI}$), and the brightness temperature of the detected sources are reported in Table~\ref{tab_detected_sources}.

In Fig.~\ref{maps} we show an example of 5\,GHz images of the FSRQ source 1FHL\,J0043.7+3425 (left panel), and the BL Lac source 1FHL\,J0648.9+1516 (right panel), which are representative of the global dataset. For these 67 sources we reveal for the first time a VLBI compact structure. The detection is also confirmed by a visual inspection of the visibility data.

\begin{figure*}
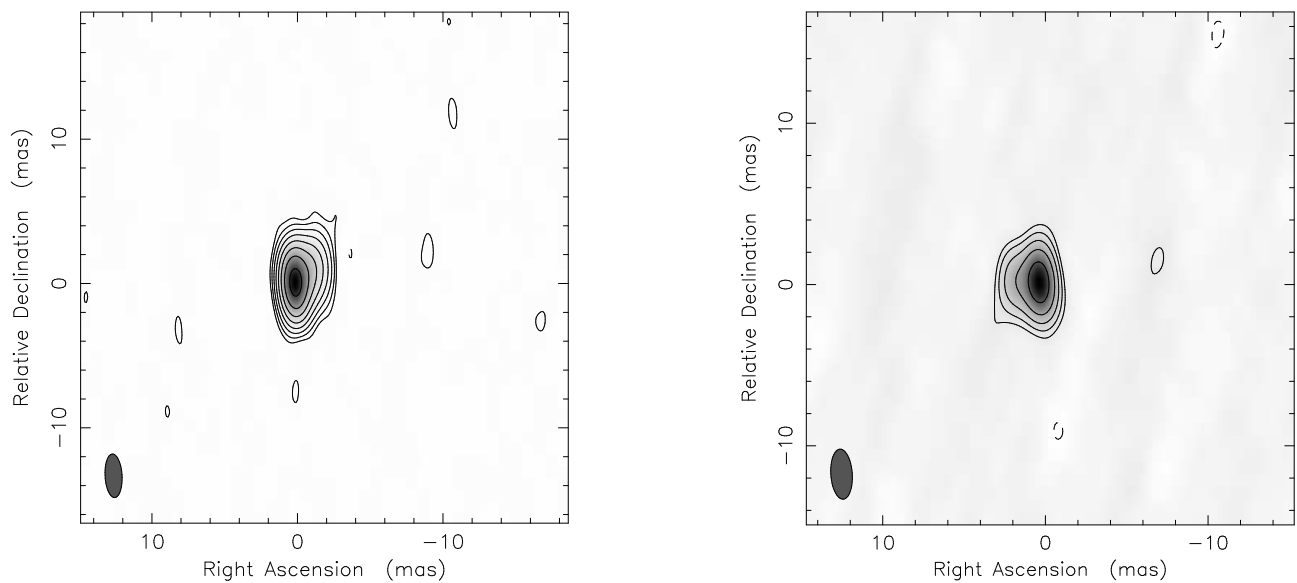

\includegraphics[bb=70 154 455 641, width=0.85\columnwidth, angle=-90, clip]{J1315.ps}
\includegraphics[bb=70 154 470 641, width=0.85\columnwidth, angle=-90 ,clip]{J1942.ps} \\
\caption{5\,GHz VLBA image of the radio counterparts of the $\gamma$-ray sources 1FHL J1315.0+2346 (left panel), classified as a BL Lac, and 1FHL J1942.8+1034 (right panel), classified as active galaxy of uncertain type. The beam size is 1.2 mas $\times$ 3.0 mas and 1.3 mas $\times$ 3.1 mas, respectively. Levels are drawn at $(-1, 1, 2, 4...) \times$ the lowest contour (that is, at 0.4 mJy/beam for the source 1FHL J1315.0+2346 and at 2.0 mJy/beam for the source 1FHL J1942.8+1034) in steps of 2. The noise level is 0.12 mJy and 0.50 mJy, respectively.} 
\label{ext_maps}
\end{figure*}

In 5 sources (1FHL J0338.4+1304, 1FHL J1315.0+2346, 1FHL J1809.3+2040, 1FHL J1942.8+1034 and 1FHL J2223.4+0104) we reveal some extended structure close to the compact region. 
Examples of extended sources are shown in Fig.~\ref{ext_maps}. 
We used the AIPS task JMFIT to fit the brightness distribution of each extended source in the image plane with two elliptical Gaussian components. In Table~\ref{ext_modelfit} we report the full widths at half maximum (FWHM) of the major and minor axes of the deconvolved elliptical Gaussian component measured in mas.

\begin{table*}
\centering
\caption{Modelfit details for the extended sources.}
\label{ext_modelfit}
\begin{tabular}{llcccccc}
\hline
\hline
1FHL name & Component & S$^{\tablefootmark{(a)}}$ & $\sigma_S$ & a$^{\tablefootmark{(b)}}$ & b$^{\tablefootmark{(b)}}$ & $D_{core}^{\tablefootmark{(c)}}$ & Orientation$^{\tablefootmark{(d)}}$\\
 & & mJy & mJy & mas & mas & mas & deg \\
\hline
 &&&&&&& \\
J0338.4+1304 & Core & 6.33  & 0.29 & 0.80 & 0.00 &   -  &  -    \\
             & C1   & 3.94  & 0.29 & 0.39 & 0.00 & 1.46 & -28.0 \\
J1315.0+2346 & Core & 51.27 & 0.12 & 0.22 & 0.00 &   -  & -       \\
             & C1   & 30.57 & 0.33 & 1.76 & 0.72 & 0.50 & -49.1 \\
J1809.3+2040 & Core & 26.82 & 0.20 & 0.67 & 0.00 &   -  & -       \\
             & C1   & 14.48 & 0.72 & 5.02 & 1.46 & 1.36 & -18.7 \\
J1942.8+1034 & Core & 53.30 & 0.78 & 0.73 & 0.52 &   -  & -     \\  
             & C1   & 13.07 & 0.41 & 1.06 & 0.00 & 0.72 &  89.0 \\
J2223.4+0104 & Core & 6.80  & 0.24 & 0.96 & 0.30 &   -  & -     \\
             & C1   & 3.41  & 0.13 & 0.88 & 0.00 & 1.35 & 157.4 \\
\hline
\end{tabular} 
\tablefoot{
\begin{small}
\newline
\tablefoottext{a}{Fitted core flux density at 5\,GHz.}\\
\tablefoottext{b}{$a$ and $b$ are the nominal FWHM of the major and minor axes of the deconvolved elliptical Gaussian component measured in mas}\\
\tablefoottext{c}{Distance from the core of the C1 Gaussian component.}\\
\tablefoottext{d}{Orientation of the extended structure. Angles are measured from North through East.}\\
\end{small}
}
\end{table*}

\begin{figure}
\includegraphics[width=0.95\columnwidth]{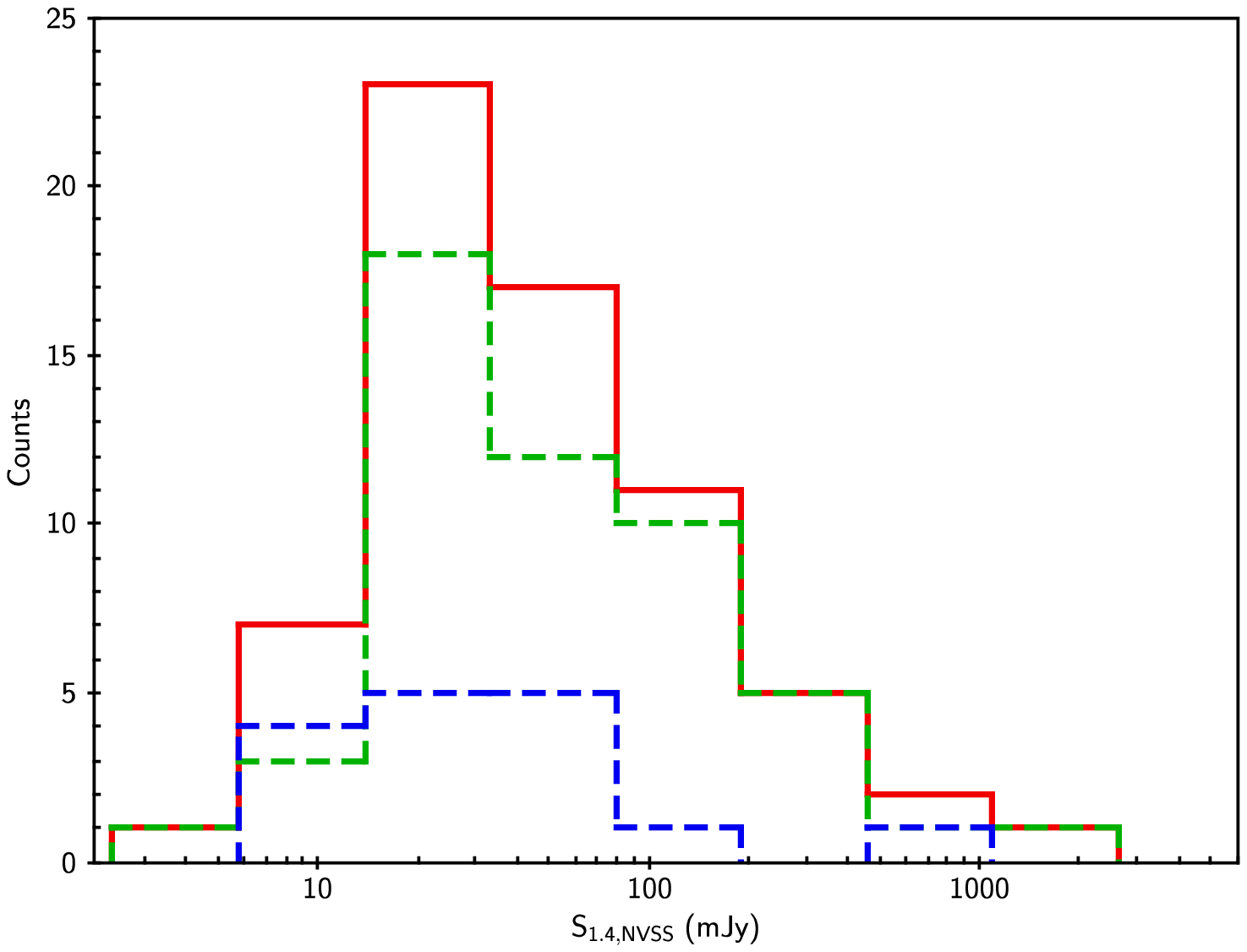} \\
\includegraphics[width=0.95\columnwidth]{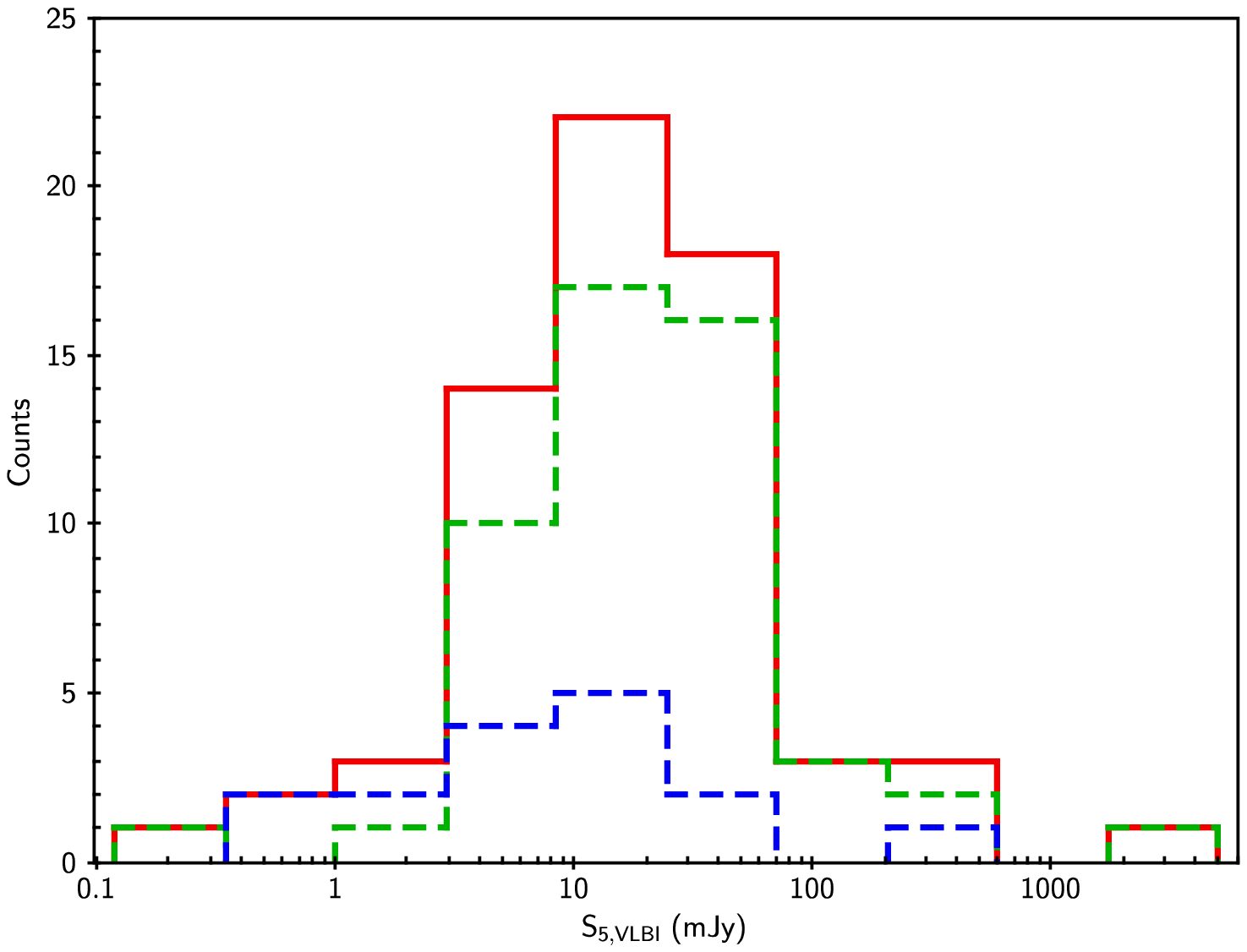} \\
\includegraphics[width=0.95\columnwidth]{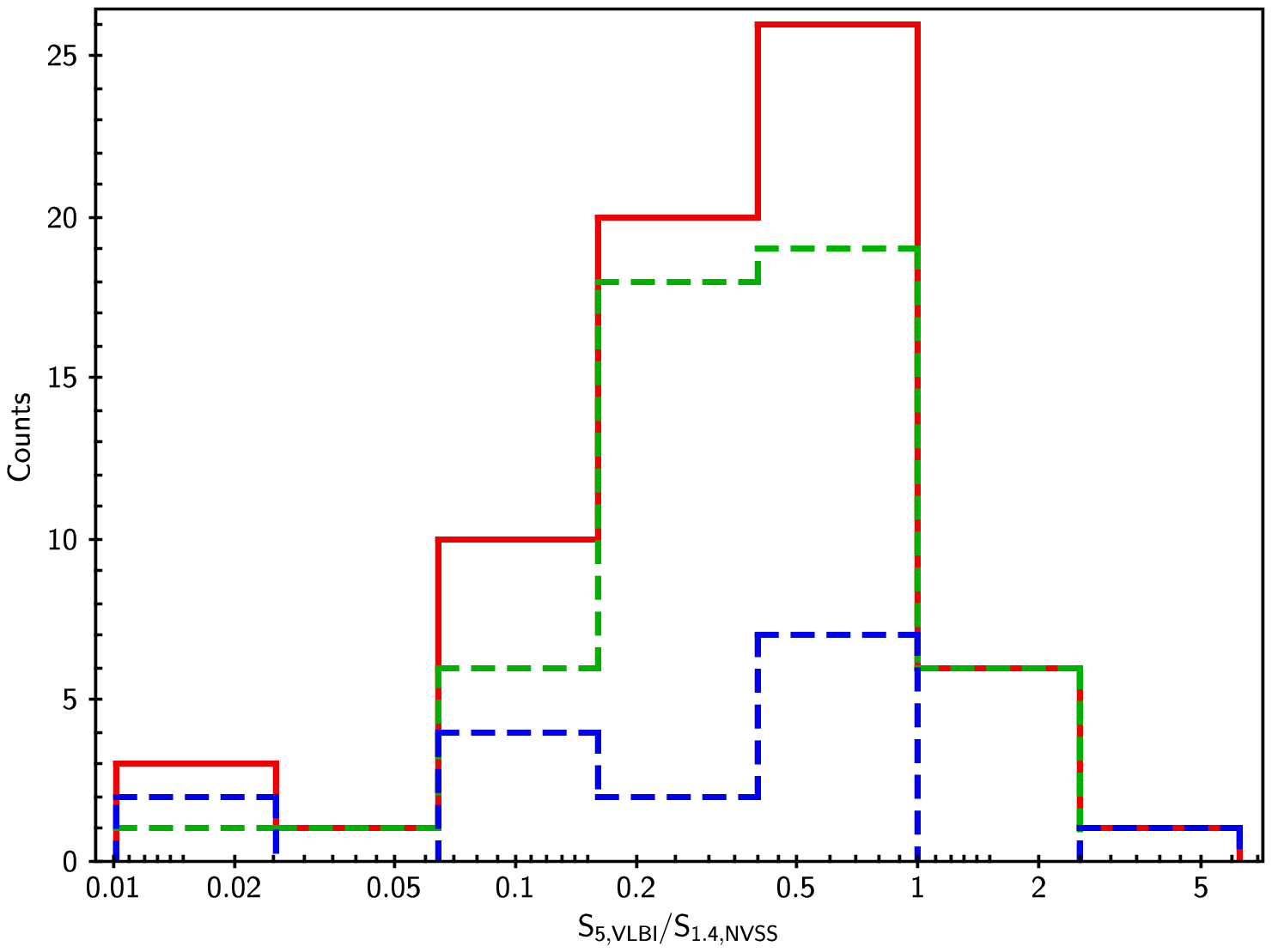} 
\caption{\small Flux density distribution for the detected sources. Top panel: 1.4\,GHz NVSS flux density distribution; middle panel: 5\,GHz VLBA flux density distribution; bottom panel: distribution of the ratio between the 5\,GHz VLBA and the 1.4\,GHz NVSS flux densities. Solid red line: all sources; green dashed line: blazars; blue dashed line: UGS.} 
\label{radio_histo_VLBA_detected}
\end{figure}

\subsection{Flux density distribution}

The distribution of $S_{1.4, NVSS}$ of the newly VLBA detected sources is shown in the top panel of Fig.~\ref{radio_histo_VLBA_detected} (red solid line), which has a median value of 40.6 mJy. In the middle panel of Fig.~\ref{radio_histo_VLBA_detected} we show the $S_{5, VLBI}$ distribution (red solid line), which has a median value of 16.3 mJy. In both panels we also show separately the distribution of flux densities for the blazars (green dashed line) and for the UGS (blue dashed line).
We note that for the UGS the median values for both $S_{1.4, NVSS}$ (31.2 mJy) and $S_{5, VLBI}$ (8.9 mJy) are lower than for the blazars (median values of 47.3 mJy and 19.5 mJy for $S_{1.4, NVSS}$ and $S_{5, VLBI}$, respectively).

In the bottom panel of Fig.~\ref{radio_histo_VLBA_detected} we show the distribution of the logarithm of the ratio between the flux density on parsec scale ($S_{5, VLBI}$) and on kilo-parsec scale ($S_{1.4, NVSS}$). The distribution peaks at $\sim0.5$, indicating that for most sources there is a considerable amount of radio emission on scales intermediate between parsec and kilo-parsec scale.

By an inspection of both NVSS and VLBI flux densities, it emerges that there are some interesting sources. In $\sim10$\% of the sources the ratio $S_{5, VLBI}/S_{1.4, NVSS}$ is significantly higher than 1, like in the case of the UGS 1FHL\,J0307.4+4915, with $S_{5, VLBI}= 300 \pm 30$ mJy, while $S_{1.4, NVSS}$ is $56.0 \pm 1.7$ mJy. 
On the other end of the distribution there are sources for which the ratio $S_{5, VLBI}/S_{1.4, NVSS}$ is significantly lower than 1. For example, for the sources 1FHL\,J0333.6+2918, 1FHL\,J0334.0+6539, 1FHL\,J0601.0+3838, and 1FHL\,J2127.8+3614, the ratio $S_{5, VLBI}/S_{1.4, NVSS}$ is between 0.1 and 0.2.

The immediate interpretation is that the sources with $S_{1.4, NVSS} \gg S_{5, VLBI}$ have extended emission (on arcsecond scale), while in the case of $S_{5, VLBI} \gg S_{1.4, NVSS}$ some degree of variability could be responsible of the higher observed VLBI flux density. 
However, it is important to stress that we are comparing non concurrent observations at different spatial scales and at different frequencies.

\subsection[High energy properties]{High energy properties}
\begin{figure*}
\includegraphics[width=0.95\columnwidth]{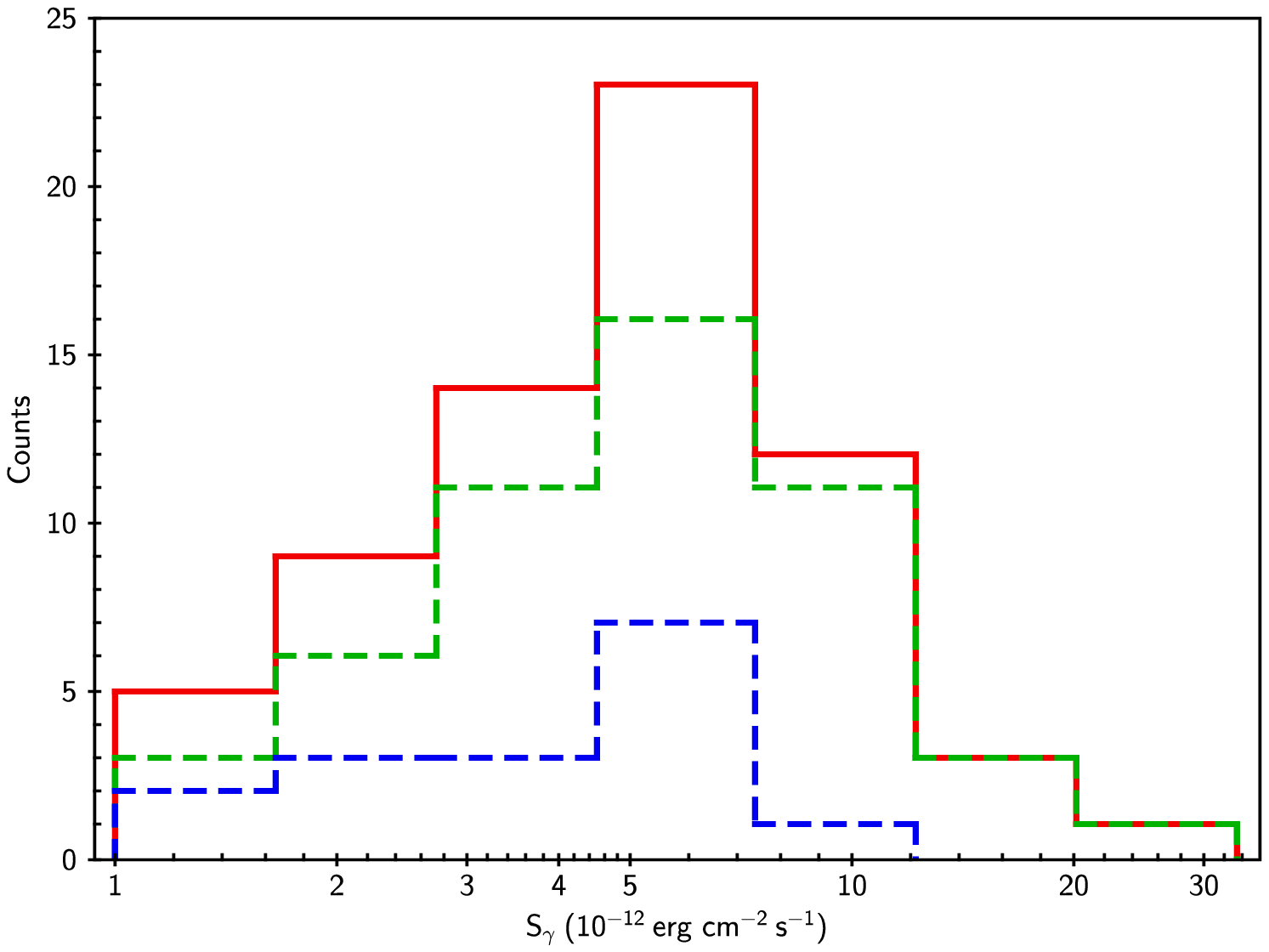} 
\includegraphics[width=0.95\columnwidth]{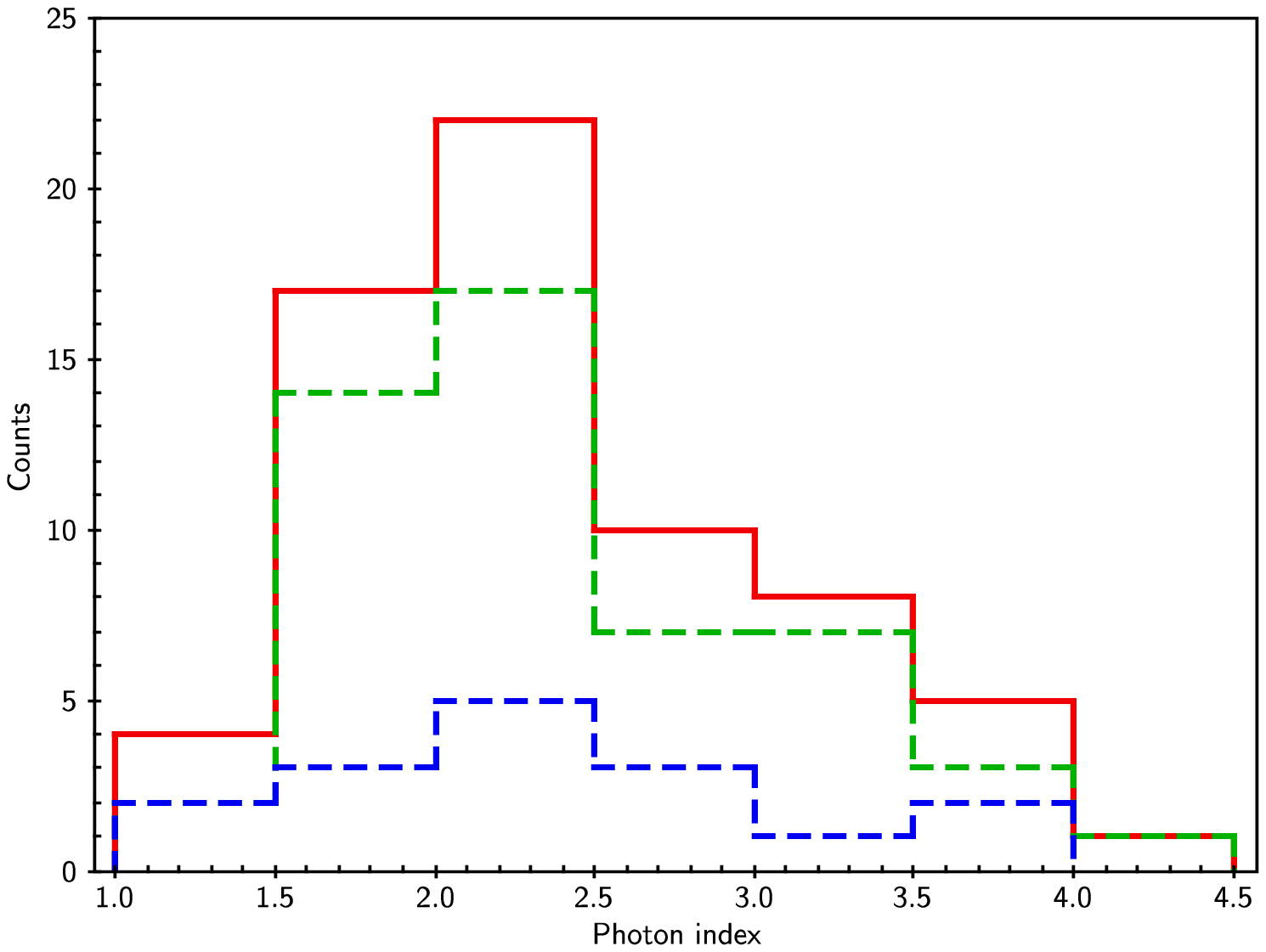} \\
\caption{\small Distribution of $\gamma$-ray energy flux (left panel) and photon index (right panel) above 10\,GeV for the detected sources. Solid red line: all sources; green dashed line: blazars; blue dashed line: UGS.} 
\label{gamma_histo_VLBA_detected}
\end{figure*}

The distribution of 1FHL $S_{\gamma}$ for the detected sources is reported in the left panel of Fig.~\ref{gamma_histo_VLBA_detected} (red solid line), showing a median value of $4.8 \times 10^{-12}$ erg\,cm$^{-2}$\,s$^{-1}$. In particular, blazars (green dashed lines) and the UGS (blue dashed lines) have median values of $5.1 \times 10^{-12}$ and $4.5 \times 10^{-12}$ erg\,cm$^{-2}$\,s$^{-1}$, respectively. 

The photon index distribution is shown in the right panel of Fig.~\ref{gamma_histo_VLBA_detected}, and has a median value of 2.3 both for the blazars (green dashed lines) and the UGS (blue dashed lines).

\subsection{Proposed counterparts for unassociated sources}
\label{sec_unassociated}
For 16 out of the 21 UGS studied in this paper, we detect radio emission on VLBI scales and we propose a possible radio NVSS counterpart. 
By using the approach proposed by \citet{Schinzel2015}, we determine the likelihood ratio for the association of each of the detected UGS.
\citet{Schinzel2015} propose a likelihood ratio threshold value of 8, chosen from the distribution of the obtained values of their full sample, to claim the association of the $\gamma$-ray source with the proposed radio counterpart. For all of the detected UGS of the present work but two (1FHL J0605.4+2726 and 1FHL J1212.2+2318) the likelihood ratio of their association is higher than 8.
The list of the detected UGS, along with further details about other $\gamma$-ray associations, is reported in Table~\ref{tab_unassoc}.

We note that 8 out of the 16 detected UGS of our sample are included in the all-sky radio survey, between 5 and 9\,GHz, of sky areas surrounding all of the the second \fermi-LAT catalog \citep[2FGL,][]{Nolan2012} UGS by \citet{Schinzel2015}. 
Our proposed low-frequency associations always match with those proposed by \citet{Schinzel2015} and by using their 7\,GHz VLBA observations we obtain the spectral index values between 5 and 7\,GHz (see Table~\ref{tab_unassoc}).

For these 8 1FHL UGS included in the 2FGL catalog, our proposed low-frequency counterparts are in agreement with those proposed in other works which make use of complementary association methods based on the analysis of their multi-wavelength (MWL) properties, such as the investigation of infrared colors and the low frequency spectral index \citep{Massaro2013a, Massaro2013b, Massaro2014, Massaro2015}.  
Our proposed counterparts for the sources 1FHL J0644.2+6036, 1FHL J1223.3+7953 and 1FHL J0601.0+3838 are compatible with the X-ray associations proposed by \citet{Landi2015} and \citet{Paggi2013}, as indicated in the notes in Table~\ref{tab_unassoc}. 

In Table~\ref{tab_unassoc}, we further report the sources 1FHL J0332.1+6307, 1FHL J0338.4+1304, 1FHL J0425.3+6320, 1FHL J0601.0+3838, 1FHL J0650.4+2056, 1FHL J1841.1+2914, 1FHL J2002.6+6303, and 1FHL J2004.7+7003, for which new low-frequency counterparts are proposed in the 3FGL catalog. For the source 1FHL J0307.4+4915 a new low-frequency counterpart is now proposed in the second catalog of hard \fermi-LAT sources \citep[2FHL,][]{Ackermann2016}. All of these new associations are coincident with those proposed in this work. 

For the sources 1FHL J0053.9+4030 and 1FHL J1619.8+7540 we propose, for the first time, a new low-frequency association. We note that the source 1FHL J0053.9+4030 is one of the faintest sources ($S_{VLBI}\sim5$ mJy) of our sample and it is not included in any other \fermi\ catalog.
For the sources 1FHL J0605.4+2726 and 1FHL J1212.2+2318, which are not included in any other \fermi\ catalog, we propose a new low-frequency counterpart, but for their association we find a likelihood ratio lower than 8. In both cases the proposed radio counterpart is located very close to the edge of the 1FHL error ellipse.

Among the 5 undetected UGS, the sources 1FHL J1406.4+1646 and 1FHL J0828.9+0902 are now associated in the 3FGL with sources not included among our observed phase centers; the source 1FHL J0625.9+0002 is included in the 3FGL and classified as UGS; the sources 1FHL J0625.9+0002 and 1FHL J0432.2+5555 are not included in any other \fermi\ catalog.

\begin{table*}
\centering
\caption{Details about the 16 detected UGS.}
\label{tab_unassoc}
\begin{tabular}{cccccccl} 
\hline
\hline
1FHL name  & NVSS assoc & 2FGL name & 3FGL name & 2FHL name & $\alpha$\tablefootmark{(a)} & $\sigma_{\alpha}$ & Notes \\
\hline
J0053.9+4030 & J005330+403015 &      -       & -            & -            &   -   & -   &  -                  \\
J0307.4+4915 & J030727+491510 & J0307.4+4915 & J0307.3+4916 & J0307.4+4917 &  -0.3 & 0.4 &  \textit{1,2,3}     \\   
J0332.1+6307 & J033153+630814 & J0332.1+6309 & J0332.0+6308 & -            &  -1.0 & 0.4 &  \textit{1,2,3}     \\   
J0338.4+1304 & J033829+130215 & J0338.2+1306 & J0338.5+1303 & -            &  -2.3 & 0.4 &  \textit{2,3}       \\   
J0425.3+6320 & J042524+632005 &      -       & J0425.2+6319 & J0425.1+6317 &  -    &  -  &  -                  \\ 
J0601.0+3838 & J060102+383828 & J0600.9+3839 & J0601.0+3837 & J0601.0+3837 &  -0.2 & 0.4 &  \textit{1,2,3,5,6} \\   
J0605.4+2726 & J060501+272456 &      -       & -            & -            &   -   & -   &  -                  \\
J0644.2+6036 & J064435+603849 & J0644.6+6034 & J0644.6+6035 & -            &  -2.7 & 0.4 &  \textit{1,3,4,5}   \\   
J0650.4+2056 & J065035+205556 &      -       & J0650.5+2055 & J0650.5+2056 &   -   & -   &  -                  \\
J1212.2+2318 & J121238+231110 &      -       & -            & -            &   -   & -   &  -                  \\
J1223.3+7953 & J122358+795329 & J1223.3+7954 & J1222.7+7952 & -            &  -0.7 & 0.4 &  \textit{1,2,3,5,7} \\   
J1548.3+1455 & J154824+145702 & J1548.3+1453 & J1548.4+1455 & -            &  -1.1 & 0.4 &  \textit{2,3,5}     \\   
J1619.8+7540 & J161913+753753 &      -       & J1619.1+7538 & -            &  -    &  -  &  -                  \\ 
J1841.1+2914 & J184121+290945 &      -       & J1841.2+2910 & -            &  -    &  -  &  -                  \\ 
J2002.6+6303 & J200244+630230 &      -       & J2002.7+6303 & -            &  -    &  -  &  -                  \\ 
J2004.7+7003 & J200506+700440 & J2004.6+7004 & J2004.8+7003 & -            &  -2.2 & 0.4 &  \textit{2,3,4,5}   \\   
\hline
\end{tabular} 
\tablefoot{
\begin{small}
\newline
\tablefoottext{a}{Spectral index calculated by using the 5\,GHz VLBA flux densities of the present work together with the 7\,GHz VLBA flux densities provided by \citet{Schinzel2015}.}\\
\tablefoottext{1}{Association also proposed by \citet{Massaro2013a}.}\\
\tablefoottext{2}{Association also proposed by \citet{Massaro2015}.}\\
\tablefoottext{3}{Association also proposed by \citet{Schinzel2015}.}\\
\tablefoottext{4}{Association also proposed by \citet{Massaro2013b}.}\\
\tablefoottext{5}{Association also proposed by \citet{Massaro2014}.} \\
\tablefoottext{6}{Association also proposed by \citet{Paggi2013}.} \\
\tablefoottext{7}{Association also proposed by \citet{Landi2015}.}
\end{small}
}
\end{table*}

\subsection{Measurements of the brightness temperature}
The high resolution of VLBI observations allows a direct measurement of the brightness temperature of a source. However, being blazar jets highly relativistic, and therefore Doppler boosted, it is difficult to measure the intrinsic brightness temperature ($T_{B}^{int}$). What we measure is the observed brightness temperature ($T_{B}^{obs}$), which is linked to $T_{B}^{int}$ by the relation $T_{B}^{obs} = \delta^{3+\alpha} \cdot T_{B}^{int}$, where $\delta$ is the Doppler factor and $\alpha$ is the spectral index.

By fitting the brightness distribution of each source in the image plane with elliptical Gaussian components, and assuming $\alpha=0$, we determine $T_B^{obs}$ for each source with the following equation \citep{Piner1999, Tingay1998}:

$$T_{B}^{obs}=1.22 \times 10^{12} \frac{S_{core}(1+z)}{ab\nu^2},$$

where $S_{core}$ corresponds to the fitted core flux density at 5\,GHz measured in Jy, $a$ and $b$ are the full widths at half maximum (FWHM) of the major and minor axes of the elliptical Gaussian core component measured in mas, $z$ is the redshift, and $\nu$ is the observing frequency in GHz.

For the sources without any redshift estimation we assume $z$ = 0.2, which corresponds to the median redshift value of 1FHL BL Lac objects.
We note that for some sources the radio core is only slightly resolved or unresolved (i.e. less than half of the beam size). 

For those sources we determine measurements as upper limits to $a$ and $b$ based on the Gaussian deconvolved size reported by the AIPS task JMFIT \citep{Greisen2003}. When a nominal value is provided we use it as measurement, when only a maximum value is reported we treat it as an upper limit, resulting in a $T_{B}^{obs}$ lower limit.
In a few cases, also the maximum value reported by JMFIT is 0; for these sources we conservatively set the upper limit to the minimum size value found among the other sources.


The resulting brightness temperature values are reported in Table~\ref{tab_detected_sources}. For the 36 sources whose radio core is resolved, the average $T_{B}^{obs}$ is of the order of 2 $\times 10^{10}$ K, which is close to the expected value for equipartition. Only five sources have $T_{B}^{obs}$ varying in the range 7 $\times 10^{10}$ K - 1 $\times 10^{11}$ K.

\section{Discussion} 
\label{sec.discussion}

\subsection{General characterization of the observed sample}
The 1FHL catalog provides us, for the first time, with a deep, large and unbiased sample of sources detected in the energy range 10-500\,GeV. It fills the gap existing between the HE and VHE regimes, represented by 3FGL and TeVCat\footnote{\url{http://tevcat.uchicago.edu/}}, respectively.

\begin{figure}
\centering
\includegraphics[width=0.95\columnwidth]{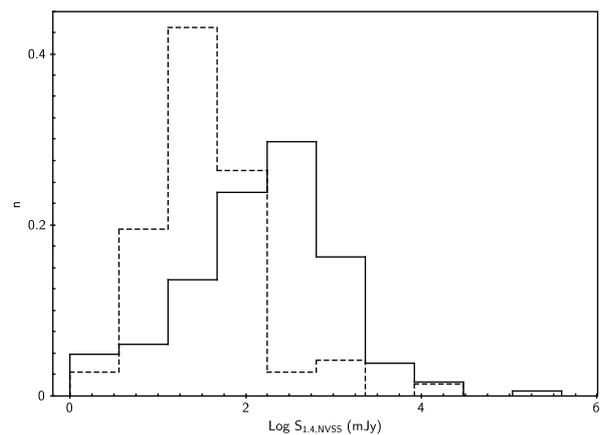}  \\
\caption{Histogram of the NVSS flux density distribution for VLBA observed sources (solid line) and unobserved sources (dashed line) of our sample.} 
\label{histo_flux_comparison}
\end{figure}

An important step on the basis of the analysis and interpretation of the physical properties of these sources is the investigation and comparison of the multi-frequency properties for the entire catalog.
However, such a study cannot be carried out neglecting a significant fraction of the sample, i.e. the 84 sources without VLBI data presented here.
These 84 sources do not represent a random 1FHL sub-sample. Their 1.4\,GHz NVSS radio flux density distribution is significantly different with respect to the other sources belonging to the 1FHL-n sample (Fig.~\ref{histo_flux_comparison}). A KS-test provides a probability of the two distributions being drawn from the same population of $\sim 2 \times 10^{-12}$.
Another peculiarity of our sample is that $61$\% of the selected AGNs are of HSP type, while in the complementary sample of 1FHL-n sources already observed with VLBI the fraction of HSPs is $40$\%.
This is mainly because we are targeting a 1FHL sub-sample of sources with the faintest radio flux densities, in agreement with the blazar sequence trend \citep[e.g.][]{Fossati1998, Ghisellini1998}.
Moreover, the fraction of HSP objects without a known redshift in our sample is $65$\%, while in the 1FHL-n they represent $42$\%. Finally, we note that while $10$\% of 1FHL sources at dec$>0^{\circ}$ are classified as UGS, in our sample the fraction of UGS is $29$\%. We are targeting the $88$\% (21/24) of the 1FHL UGS in the northern sky with at least one NVSS counterpart within the r95. 
Therefore, we are not only completing the VLBI data collection for the whole sample of hard $\gamma$-ray sources but we are characterizing an unexplored population.

\subsection{VLBI properties and unassociated sources}
The vast majority of the detected sources of our sample are radio weak (with flux densities of few mJy).  
For all of the detected sources we reveal a VLBI compact component and we confirm their blazar nature. This finding provides an independent validation of other association methods based on the analysis of their multi-wavelength properties, such as the investigation of infrared colors, the low frequency spectral index and the X-ray association \citep[e.g.,][]{D'Abrusco2013, Massaro2013a, Nori2014, Landi2015}.
 
It is interesting to note that the detection rate of compact components for the UGS is remarkably high (76\%). This is a first indication that the nature of the UGS is similar to the associated sources, i.e. they are mostly blazars. This hypothesis is supported from the fact that both $S_{1.4, NVSS}$ and $S_{5, VLBI}$ for the blazars and the UGS, shown in Fig.~\ref{radio_histo_VLBA_detected}, are distributed in a similar way. This also shows how much the NVSS survey is useful for a pre-selection of compact radio counterparts, and deeper Jansky very large array and Australia telescope compact array observations are only needed in special cases \citep{Schinzel2015}. However, the VLBI observations are necessary for the actual selection of the right association.

We note that for the UGS both the NVSS and VLBI flux densities are lower than for the associated ones (Fig.~\ref{radio_histo_VLBA_detected}), therefore they may be weaker blazars not yet identified.
Moreover, for eight detected UGS we calculated the spectral index values between 5 and 7\,GHz VLBA data, which are either flat or inverted, indicating the presence of a self-absorbed region. In particular for the source 1FHL J0644.2+6036 the spectral index is strongly inverted ($\alpha$=-2.7), exceeding the canonical value of -2.5 for a homogeneous synchrotron emission region. This is an indication that variability is occurring in this source. A clear example of flux density variability is provided by the source 1FHL J0307.4+4915, for which we find $S_{5, VLBI}= 300 \pm 30$ mJy, while $S_{1.4, NVSS}$ is $56.0 \pm 1.7$ mJy.

\subsection{Brightness temperature}
For the detected sources we estimated the VLBI core brightness temperature, which can be used to discern among the physical processes occurring in the source.
In a synchrotron source there are two main physical mechanisms which can limit the intrinsic brightness temperature. One is the inverse Compton (IC) process which limits $T_{B}^{int}$ in the range $\sim 5 \times 10^{11} - 10^{12}$ K \citep{Kellermann1969}. When this limit is exceeded the IC process leads to rapid and catastrophic electron energy losses. The second relevant physical mechanism is the energy equipartition between particles and magnetic field which limits the $T_{B}^{int}$ in the range $\sim 5 \times 10^{10} - 10^{11}$ K \citep{Readhead1994}.

For the vast majority of the sources of our sample whose core is resolved we obtain brightness temperature values of the order of $2\times 10^{10}$ K, which is close to the expected value for equipartition. Therefore, what we expect is that the energy is almost equally stored in the magnetic field and in the radiating particles.
These values are consistent with the typical HSP blazar $T_{B}^{obs}$, and are in agreement with the values obtained in other works \citep[e.g.][]{Piner2010, Lico2012, Lico2014, Piner2014}.
From these $T_{B}^{obs}$ values no high Doppler factors are required and there is no evidence of a strong beaming. However, we note that at 5\,GHz we are observing these sources in their self-absorbed part of the synchrotron spectrum, therefore the flux densities and the brightness temperatures may be underestimated.

\section{Summary and conclusions}
\label{sec.conclusions}
In the present work we targeted and characterized the 1FHL sources in the northern sky with the faintest radio flux densities, which represent an important extension to an unexplored region of the parameter space, and are essential to gather a sample as large and unbiased as possible to explore the possible correlation between radio VLBI and E >10\,GeV emission.

\begin{itemize}

\item 
For all of the detected sources we reveal a compact high brightness temperature VLBI radio core and we confirm their blazar nature. The vast majority of them are radio weak sources, with a VLBI flux density median value of 16.3 mJy, consistently with the predictions of the blazar sequence \citep[e.g.][]{Fossati1998, Ghisellini1998}.

\item 
Thanks to the new VLBI observations we propose new low-frequency counterparts for 16 1FHL UGS. We note that for 12 out of the 16 detected UGS our proposed low-frequency counterparts are in agreement with those proposed by using indirect and complementary association methods based on the analysis of their MWL properties, and with those included in the 2FHL and 3FGL catalogs.
For the remaining four 1FHL UGS we propose for the first time a low-frequency counterpart.

\item 
For the sources whose radio core is resolved we obtain brightness temperature values of the order of $2 \times 10^{10}$ K, which are close to the value expected for the equipartition of energy between particles and magnetic field. Therefore, there is no evidence of a strong beaming and no high Doppler factor values are required.
This result increases the statistical significance of the similar brightness temperature values obtained for other HSP sources \citep[e.g. ][]{Giroletti2004, Piner2014}.
\end{itemize}

While for blazars belonging to the first catalog of AGNs detected by \fermi\ \citep[1LAC, ][]{Abdo2010b}, \citet{Ackermann2011} found a strong and significant correlation between radio and $\gamma$ rays in the 0.1-100\,GeV energy range, no systematic investigation for the existence of a possible correlation between radio and VHE $\gamma$ rays has been made. The main reason is that at VHE there is no homogeneous, deep and large survey which allows us to perform a similar correlation analysis. This is because the IACTs have small field of view and they mainly observe in pointing mode. All of these limitations will be overcome in the next years thanks to the new generation Cherenkov Telescope Array \citep[CTA, ][]{Dubus2013}.  
The investigation in the VHE domain is important because it provides us with details about the blazar sequence (i.e. the anti-correlation between the synchrotron luminosity and the SED peak frequency \citep{Fossati1998}), and the interaction of VHE photons with the EBL \citep[e.g. ][]{Ackermann2012}.

In a following paper, by using the VLBI flux densities presented in this work complemented by VLBI archival observations, we will explore and quantify the possible radio-VHE correlation for the 1FHL AGN sample.

\begin{acknowledgement}
\begin{small}
We thank Leonid Petrov for his support and the fruitful discussions. We thank the anonymous referee for the helpful comments and discussions which improved the manuscript.
This work is based on observations obtained through the S6340 VLBA project, in the framework of the \fermi -NRAO cooperative agreement. The VLBA makes use of the Swinburne University of Technology software correlator, developed as part of the Australian Major National Research Facilities Programme and operated under licence \citep{Deller2011}. We acknowledge financial contribution from grant PRIN-INAF-2011 and PRIN-INAF-2014. 
This research has made use of NASA's Astrophysics Data System, of the VizieR catalog access tool, CDS, Strasbourg, France and the TOPCAT software \citep{Taylor2005}.
The National Radio Astronomy Observatory is a facility of the National Science Foundation operated under cooperative agreement by Associated Universities, Inc. For this paper we made use of the NASA/IPAC Extragalactic Database NED which is operated by the JPL, Californian Institute of Technology, under contract with the National Aeronautics and Space Administration. 
The \textit{Fermi}-LAT Collaboration acknowledges support for LAT development, operation and data analysis from NASA and DOE (United States), CEA/Irfu and IN2P3/CNRS (France), ASI and INFN (Italy), MEXT,
KEK, and JAXA (Japan), and the K.A.~Wallenberg Foundation, the Swedish Research Council and the National Space Board (Sweden). Science analysis support in the operations phase from INAF (Italy) and CNES (France) is also gratefully acknowledged.

\end{small}
\end{acknowledgement}

\addtocounter{table}{-5}

\onecolumn

\setlength{\tabcolsep}{2.5pt}
\longtab{}{
\begin{small}
\begin{longtable}{lccccccccc}
\caption{\label{tab_sources_log} Sample sources details. Col.\,1: 1FHL name; col.\,2: low frequency association of the 1FHL source; col.\,3: source classification: bzb = blazar of the BL Lac type, bzq = blazar of the FSRQ type, agu = active galaxy of uncertain type, snr = supernova remnant; col.\,4: redshift; cols.\,5 and 6: pointing coordinates; cols.\,7 and 8: R.A. and Dec. shift from the phase tracking center; col.\,9: detection; col.\,10: Experiment code.}\\
\hline\hline
\setlength{\tabcolsep}{2.5pt}
IFHL name\tablefootmark{a} & Assoc & Class & z & R.A.\tablefootmark{b} & Dec\tablefootmark{b} & Shift R.A. & Shift Dec & Detection & Exp. code \\
 & & & & h:m:s & d:m:s.& mas & mas & & \\
 \hline
\endfirsthead
\caption{Continued.}\\
\hline\hline
IFHL name\tablefootmark{a} & Assoc & Class & z & R.A.\tablefootmark{b} & Dec\tablefootmark{b} & Shift R.A. & Shift Dec & Detection & Exp. code \\
 & & & & h:m:s & d:m:s.& mas & mas & & \\
\hline
\endhead
\hline
\endfoot
 & & & & & & & & & \\
J0007.7+4709	&	MG4 J000800+4712	&	bzb	&	0.28&	00:07:45.22	&	+47:11:31.0	&	-135.6	&	31.4	&	Y	&	S6340H	\\
J0009.2+5032	&	NVSS J000922+503028	&	bzb	&	-	&	00:09:22.53	&	+50:30:28.9	&	-2190.5	&	52.0	&	Y	&	S6340A	\\
J0037.8+1238	&	NVSS J003750+123818	&	bzb	&	0.09&	00:37:50.85	&	+12:38:18.8	&	-443.0	&	-1079.5	&	Y	&	S6340A	\\
J0040.3+4049	&	1ES 0037+405	    &	bzb	&	-	&	00:40:13.50	&	+40:50:05.4	&	-3552.4	&	704.4	&	Y	&	S6340B	\\
J0043.7+3425	&	GB6 J0043+3426	    &	bzq	&	0.97&	00:43:48.86	&	+34:26:26.3	&	210.0	&	206.5	&	Y	&	S6340A	\\
J0053.9+4030	&	UGS          	    &	-	&	-	&	00:53:31.76	&	+40:27:21.5	&	-867.8	&	-3859.8	&	Y	&	S6340H	\\
J0103.4+5336	&	1RXS J010325.9+533721&	agu	&	-	&	01:03:26.08	&	+53:37:12.1	&	1085.0	&	-1179.0	&	Y	&	S6340B	\\
J0122.7+3425$^*$&	1ES 0120+340	    &	bzb	&	0.27&	01:22:38.81	&	+34:26:11.3	&	-    	&	-	    &	N	&	S6340B	\\
J0131.3+6121	&	1RXS J013106.4+612035&	agu	&	-	&	01:31:07.16	&	+61:20:33.0	&	-478.5	&	-368.0	&	Y	&	S6340B	\\
J0134.4+2643	&	1RXS J013427.2+263846&	agu	&	-	&	01:34:27.99	&	+26:38:42.5	&	-2759.0	&	-525.0	&	Y	&	S6340B	\\
J0153.1+7515	&	BZB J0153+7517	    &	bzb	&	-	&	01:53:07.26	&	+75:17:44.2	&	-459.5	&	1268.0	&	Y	&	S6340B	\\
J0213.1+2246	&	MG3 J021252+2246	&	bzb	&	0.46&	02:12:52.78	&	+22:44:55.1	&	-752.0	&	2858.0	&	Y	&	S6340B	\\
J0241.3+6548	&	NVSS J024121+654311	&	agu	&	-	&	02:41:26.41	&	+65:45:47.7	&	-471.5	&	321.0	&	Y	&	S6340H	\\
J0307.4+4915	&	UGS          	    &	-	&	-	&	03:07:27.06	&	+49:15:10.3	&	246.0	&	546.0	&	Y	&	S6340B	\\
J0322.1+2337	&	MG3 J032201+2336	&	bzb	&	-	&	03:22:00.05	&	+23:36:10.0	&	1093.5	&	-1205.5	&	Y	&	S6340B	\\
J0332.1+6307	&	UGS          	    &	-	&	-	&	03:31:53.76	&	+63:08:14.3	&	-1056.5	&	139.0	&	Y	&	S6340A	\\
J0333.6+2918	&	TXS 0330+291	    &	bzb	&	-	&	03:33:49.02	&	+29:16:31.6	&	214.0	&	51.0	&	Y	&	S6340A	\\
J0334.0+6539	&	TXS 0329+654	    &	bzb	&	-	&	03:33:56.74	&	+65:36:56.2	&	1.0	    &	-1.5	&	Y	&	S6340A	\\
J0338.4+1304	&	UGS          	    &	-	&	-	&	03:38:29.30	&	+13:02:15.0	&	305.0	&	-449.0	&	Y	&	S6340A	\\
J0425.3+6320	&	UGS          	    &	-	&	-	&	04:25:24.77	&	+63:20:05.6	&	-3944.0	 &	3918.5  &	Y	&	S6340B	\\
J0425.4+5601	&	UGS          	    &	-	&	-	&	04:26:04.31	&	+56:03:39.6	&	-	    &	-	    &	N	&	S6340B	\\
J0432.2+5555	&	UGS          	    &	-	&	-	&	04:32:02.39	&	+55:48:44.2	&	-	    &	-	    &	N	&	S6340H	\\
J0515.9+1528	&	GB6 J0515+1527	    &	bzb	&	-	&	05:15:47.41	&	+15:27:16.9	&	820.0	&	307.5	&	Y	&	S6340C	\\
J0516.4+7351	&	GB6 J0516+7350	    &	bzb	&	0.25&	05:16:31.12	&	+73:51:08.8	&	-469.0	&	140.0	&	Y	&	S6340C	\\
J0529.0+0937	&	GB6 J0529+0934	    &	agu	&	-	&	05:29:02.57	&	+09:34:32.5	&	-724.5	&	4318.0	&	Y	&	S6340C	\\
J0540.5+5822	&	GB6 J0540+5823	    &	bzb	&	-	&	05:40:29.97	&	+58:23:39.5	&	-309.0	&	1048.0	&	Y	&	S6340C	\\
J0601.0+3838	&	UGS          	    &	-	&	-	&	06:01:02.90	&	+38:38:28.8	&	362.0	&	-370.5	&	Y	&	S6340C	\\
J0605.0+0001	&	GB6 J0604+0000	    &	agu	&	-	&	06:04:58.43	&	+00:00:43.1	&	177.5	&	-254.5	&	Y	&	S6340C	\\
J0605.4+2726	&	UGS          	    &	-	&	-	&	06:05:01.05	&	+27:24:56.3	&	-1216.5	&	-1138.0	&	Y	&	S6340C	\\
J0606.6+4742$^*$&	CGRaBS J0607+4739	&	bzb	&	-	&	06:06:38.31	&	+47:41:47.4	&	-	    &	-	    &	N	&	S6340C	\\
J0625.9+0002	&	UGS          	    &	-	&	-	&	06:26:12.88	&	+00:03:34.6	&	-	    &	-	    &	N	&	S6340C	\\
J0644.2+6036	&	UGS          	    &	-	&	-	&	06:44:35.64	&	+60:38:49.8	&	-777.0	&	-1363.0	&	Y	&	S6340C	\\
J0648.9+1516	&	VER J0648+152	    &	bzb	&	0.18&	06:48:47.63	&	+15:16:25.0	&	-268.5	&	200.5	&	Y	&	S6340C	\\
J0650.4+2056	&	UGS          	    &	-	&	-	&	06:50:35.23	&	+20:55:56.8	&	-1715.0	&	-576.5	&	Y	&	S6340C	\\
J0706.5+3744	&	GB6 J0706+3744	    &	bzb	&	-	&	07:06:31.67	&	+37:44:36.7	&	-319.5	&	293.2	&	Y	&	S6340C	\\
J0743.0+5446$^*$&	GB6 J0742+5444	    &	bzq	&	0.72&	07:42:46.10	&	+54:47:06.0	&	-	    &	-	    &	N	&	S6340C	\\
J0745.2+7439	&	MS 0737.9+7441	    &	bzb	&	0.32&	07:44:05.45	&	+74:33:58.1	&	298.5	&	-155.0	&	Y	&	S6340E	\\
J0745.2+8511	&	NVSS J074715+851208	&	agu	&	-	&	07:47:15.44	&	+85:12:08.2	&	-891.0	&	-422.0	&	Y	&	S6340E	\\
J0828.9+0902	&	UGS          	    &	-	&	-	&	08:28:47.28	&	+09:03:37.3	&	-	    &	-	    &	N	&	S6340D	\\
J0850.0+4849	&	GB6 J0850+4855    	&	bzb	&	-	&	08:50:26.79	&	+48:52:00.7	&	873.1	&	-71.3	&	Y	&	S6340G	\\
J0912.5+2758$^*$&	1RXS J091211.9+275955&	bzb	&	-	&	09:12:20.81	&	+27:55:47.1	&	-	    &	-	    &	N	&	S6340D	\\
J0930.4+8611	&	S5 0916+864	        &	bzb	&	-	&	09:29:43.06	&	+86:12:21.3	&	-10.5	&	-18.5	&	Y	&	S6340E	\\
J0946.2+0106$^*$&	RXS J094620.5+010459&	bzb	&	0.56&	09:46:19.81	&	+01:06:26.1	&	-    	&	-     	&	N	&	S6340G	\\
J0959.5+6535	&	S4 0954+65	        &	bzb	&	0.37&	09:59:53.60	&	+65:34:57.0	&	-156.4	&	-518.1	&	Y	&	S6340G	\\
J1023.6+2959$^*$&	RX J1023.6+3001	    &	bzb	&	0.43&	10:23:54.58	&	+29:58:00.4	&	-	    &	-	    &	N	&	S6340D	\\
J1100.6+4018	&	RX J1100.3+4019	    &	bzb	&	0.23&	11:00:21.09	&	+40:19:27.7	&	274.5	&	-370.5	&	Y	&	S6340D	\\
J1107.5+0223	&	BZB J1107+0222	    &	bzb	&	-	&	11:07:35.99	&	+02:22:25.6	&	1112.5	&	1004.5	&	Y	&	S6340D	\\
J1130.4+5814	&	BZB J1131+5809	    &	bzb	&	0.36&	11:31:18.69	&	+58:08:58.5	&	444.0	&	-302.0	&	Y	&	S6340D	\\
J1135.7+6736$^*$&	RX J1136.5+6737	    &	bzb	&	0.13&	11:35:52.30	&	+67:37:24.2	&	-	    &	-	    &	N	&	S6340E	\\
J1137.0+2553	&	RX J1136.8+2551	    &	bzb	&	0.16&	11:37:01.13	&	+25:51:33.8	&	-1320.5	&	-1603.5	&	Y	&	S6340G	\\
J1212.2+2318	&	UGS          	    &	-	&	-	&	12:12:26.70	&	+23:14:18.9	&	487.9	&	-126.9	&	Y	&	S6340G	\\
J1223.3+7953	&	UGS          	    &	-	&	-	&	12:23:58.22	&	+79:53:29.0	&	377.5	&	781.0	&	Y	&	S6340E	\\
J1224.5+2437	&	MS 1221.8+2452	    &	bzb	&	0.22&	12:24:24.20	&	+24:36:23.6	&	134.0	&	60.5	&	Y	&	S6340E	\\
J1244.9+5708	&	1RXS J124510.5+571020&	bzb	&	-	&	12:45:09.99	&	+57:09:54.4	&	-0.5	&	-0.5	&	Y	&	S6340E	\\
J1249.7+3706	&	RX J1249.8+3708	    &	bzb	&	-	&	12:49:46.76	&	+37:07:48.0	&	140.0	&	88.0	&	Y	&	S6340E	\\
J1310.9+0036	&	RX J1311.1+0035	    &	bzb	&	-	&	13:11:06.42	&	+00:35:09.8	&	-841.0	&	-248.5	&	Y	&	S6340E	\\
J1315.0+2346	&	TXS 1312+240	    &	bzb	&	-	&	13:14:43.81	&	+23:48:26.8	&	-	    &	0.5	    &	Y	&	S6340D	\\
J1322.9+2942$^*$&	4C +29.48	        &	agu	&	-	&	13:23:02.59	&	+29:41:33.4	&	-	    &	-	    &	N	&	S6340D	\\
J1406.4+1646	&	UGS          	    &	-	&	-	&	14:06:14.40	&	+16:46:35.0	&	-	    &	-	    &	N	&	S6340E	\\
J1418.6+2539	&	BZB J1417+2543	    &	bzb	&	0.24&	14:17:56.27	&	+25:43:23.1	&	-5414.0	&	-3133.0	&	Y	&	S6340D	\\
J1419.0+7730$^*$&	1RXS J141901.8+773229&	agu	&	-	&	14:19:41.54	&	+77:34:43.6	&	-	    &	-	    &	N	&	S6340E	\\
J1548.3+1455	&	UGS          	    &	-	&	-	&	15:48:24.34	&	+14:57:02.9	&	-669.5	&	184.5	&	Y	&	S6340D	\\
J1619.8+7540	&	UGS          	    &	-	&	-	&	16:19:13.77	&	+75:37:53.8	&	100.0	&	347.5	&	Y	&	S6340E	\\
J1631.0+5224	&	TXS 1629+524	    &	bzb	&	-	&	16:30:43.12	&	+52:21:38.7	&	-261.5	&	105.5	&	Y	&	S6340D	\\
J1735.9+2033	&	NVSS J173605+203301	&	bzb	&	-	&	17:35:41.48	&	+20:35:08.8	&	821.5	&	-193.0	&	Y	&	S6340H	\\
J1744.2+1938	&	1ES 1741+196	    &	bzb	&	0.08&	17:44:15.87	&	+19:39:02.1	&	1797.8	&	-417.3	&	Y	&	S6340H	\\
J1809.3+2040	&	RX J1809.3+2041	    &	bzb	&	-	&	18:09:25.48	&	+20:41:31.0	&	394.0	&	101.1	&	Y	&	S6340F	\\
J1841.1+2914	&	UGS          	    &	-	&	-	&	18:41:20.50	&	+29:11:28.5	&	130.2	&	-29.0	&	Y	&	S6340H	\\
J1841.8+3219	&	RX J1841.7+3218	    &	bzb	&	-	&	18:41:47.17	&	+32:18:38.6	&	1625.5	&	-555.0	&	Y	&	S6340F	\\
J1911.0+0905	&	SNR G043.3-00.2	    &	snr	&	-	&	19:11:03.97	&	+09:05:02.5	&	-	    &	-	    &	N	&	S6340H	\\
J1926.9+6153	&	1RXS J192649.5+615445&	bzb	&	-	&	19:26:49.95	&	+61:54:41.4	&	414.0	&	-946.5	&	Y	&	S6340F	\\
J1942.8+1034	&	1RXS J194246.3+103339&	agu	&	-	&	19:42:47.48	&	+10:33:27.8	&	-60.0	&	698.5	&	Y	&	S6340F	\\
J2002.6+6303	&	UGS          	    &	-	&	-	&	20:02:44.96	&	+63:02:30.8	&	-2633.0 &	-2343.0 &	Y	&	S6340F	\\
J2004.7+7003	&	UGS          	    &	-	&	-	&	20:04:41.67	&	+70:06:16.8	&	1464.7	&	1340.9	&	Y	&	S6340H	\\
J2015.8+3710	&	VER J2016+372	    &	bzq	&	0.86&	20:15:53.68	&	+37:11:30.2	&	2.2	&	384.6	    &	Y	&	S6340H	\\
J2031.4+1941$^*$&	RX J2030.8+1935	    &	agu	&	-	&	20:31:40.96	&	+19:40:34.0	&	-	    &	-	    &	N	&	S6340H	\\
J2127.8+3614	&	B2 2125+35	        &	bzb	&	-	&	21:27:43.04	&	+36:13:05.0	&	138.5	&	-736.0	&	Y	&	S6340F	\\
J2212.5+2803	&	CRATES J221238+275944&	agu	&	-	&	22:12:39.10	&	+27:59:38.5	&	0.5	    &	4.5	    &	Y	&	S6340F	\\
J2223.4+0104	&	NVSS J222329+010226	&	bzb	&	-	&	22:23:29.59	&	+01:02:26.7	&	286.0	&	50.0	&	Y	&	S6340A	\\
J2247.7+4412	&	NVSS J224753+441317	&	bzb	&	-	&	22:47:53.10	&	+44:13:17.0	&	-1133.0	&	1604.0	&	Y	&	S6340A	\\
J2308.1+1459$^*$&	MG1 J230734+1449	&	bzb	&	0.50&	23:07:59.07	&	+14:58:58.4	&	-	    &	-	    &	N	&	S6340H	\\
J2314.0+1446	&	RGB J2313+147	    &	bzb	&	0.16&	23:13:57.27	&	+14:44:22.7	&	-918.5	&	-355.0	&	Y	&	S6340A	\\
J2329.1+3754	&	NVSS J232914+375414	&	bzb	&	0.26&	23:29:14.25	&	+37:54:14.9	&	-206.5	&	417.5	&	Y	&	S6340A	\\
J2347.2+0707	&	CRATES J234639+07050&	agu	&	-	&	23:47:06.48	&	+07:03:51.9	&	-1985.2	&	-2556.5	&	Y	&	S6340H	\\
\end{longtable}
\tablefoot{
\begin{tiny}
\newline
\tablefoottext{a}{With asterisks we indicate the 8 sources excluded from the analysis because of wrong pointing.}\\
\tablefoottext{b}{Coordinates of pointing.}
\end{tiny}
}
\end{small}
}

\addtocounter{table}{1}

\setlength{\tabcolsep}{3.5pt}
\begin{longtab}{}
\begin{longtable}{lcccccccccc}
\caption{\label{tab_detected_sources} Details about the 67 detected sources. Col.\,1: source 1FHL name; cols.\,2 and 3: post-shift VLBI coordinates; col.\,4: source classification: bzb = blazar of the BL Lac type, bzq = blazar of the FSRQ type, agu = active galaxy of uncertain type, UGS = unassociated $\gamma$-ray source; cols.\,5 and 6: the 10-500\,GeV energy flux of the $\gamma$-ray source and the relative 1-sigma uncertainty; cols.\,7 and 8: 1.4\,GHz NVSS flux density and the relative uncertainty; cols.\,9 and 10: 5\,GHz VLBA peak flux density and the relative uncertainty (calculated by considering a calibration error of about $15\%$ of the flux density and a statistical error provided by the map rms noise); col.\,10: brightness temperature.}\\
\hline\hline
\setlength{\tabcolsep}{3.5pt}
IFHL name & R.A. & Dec & Class & $S_{\gamma}$ & $\sigma_{S_{\gamma}}$ & $S_{NVSS}$ & $\sigma_{S_{NVSS}}$ & $S_{VLBI}$ & $\sigma_{S_{VLBI}}$ & $T_B^{\tablefootmark{a}}$ \\
 & h:m:s & d:m:s & & $10^{-12} \, erg/cm^2/s$ & $10^{-12} \, erg/cm^2/s $ & mJy & mJy & mJy & mJy & $10^9$ K \\
\hline
\endfirsthead
\caption{Continued.}\\
\hline\hline
IFHL name & R.A. & Dec & Class & $S_{\gamma}$ & $\sigma_{S_{\gamma}}$ & $S_{NVSS}$ & $\sigma_{S_{NVSS}}$ & $S_{VLBI}$ & $\sigma_{S_{VLBI}}$ & $T_B^{\tablefootmark{a}}$ \\
 & h:m:s & d:m:s & & $10^{-12} \, erg/cm^2/s$ & $10^{-12} \, erg/cm^2/s $ & mJy & mJy & mJy & mJy & $10^9$ K \\
\hline
\endhead
\hline
\endfoot
 & & & & & & & & & & \\
J0007.7+4709	&	00:07:59.973	&	47:12:07.769	&bzb&	3.8     &	1.3	    &	60.3	&	1.9	&	22.9	&	3.4	&	1.3	    \\
J0009.2+5032	&	00:09:22.760	&	50:30:28.847	&bzb&	12.3	&	3.8	    &	11.8	&	0.5	&	 5.8	&	0.9	&	0.5	    \\
J0037.8+1238	&	00:37:50.880	&	12:38:19.879	&bzb&	2.1	    &	1.2  	&	75.1	&	2.3	&	28.5	&	4.3	&	5.8	    \\
J0040.3+4049	&	00:40:13.978	&	40:50:07.145	&bzb&	8.8	    &	5.6 	&	47.3	&	1.8	&	 1.5	&	0.2	&	>0.2	\\
J0043.7+3425	&	00:43:48.843	&	34:26:26.093	&bzq&	7.6	    &	3.3 	&	92.8	&	2.8	&	106 	&	16  &	80.5	\\
J0053.9+4030	&	00:53:31.696	&	40:31:25.957	&UGS&	7.2  	&	5.4 	&	 9.6	&	0.5	&	 9.1	&	1.4	&	1.6	    \\
J0103.4+5336	&	01:03:25.958	&	53:37:13.279	&agu&	4.8 	&	1.9 	&	30.4	&	1.0	&	27.3	&	4.1	&	>32.3	\\
J0131.3+6121	&	01:31:07.227	&	61:20:33.368	&agu&	19.5	&	5.4 	&	19.1	&	0.7	&	 6.2	&	0.9	&	>11.9	\\
J0134.4+2643	&	01:34:28.196	&	26:38:43.025	&agu&	2.9  	&	1.2 	&	30.1	&	1.0	&	11.7	&	1.8	&	>16.9	\\
J0153.1+7515	&	01:53:07.380	&	75:17:42.933	&bzb&	1.6 	&	1.0 	&	21.2	&	0.8	&	 6.7	&	1.0	&	>7.8	\\
J0213.1+2246	&	02:12:52.834	&	22:44:52.242	&bzb&	5.8 	&	3.7 	&	66.9	&	2.0	&	63.0	&	9.5	&	>61.6	\\
J0241.3+6548	&	02:41:21.746	&	65:43:11.579	&agu&	4.9  	&	2.3 	&	190.3	&	5.7	&	24.0	&	3.6	&	1.6	    \\
J0307.4+4915	&	03:07:27.035	&	49:15:09.753	&UGS&	6.3  	&	3.1 	&	56.0	&	1.7	&	300 	&	45  &   104	    \\
J0322.1+2337	&	03:21:59.970	&	23:36:11.206	&bzb&	9.1 	&	4.6 	&	76.2	&	2.3	&	28.5	&	4.3	&	>21.7	\\
J0332.1+6307	&	03:31:53.916	&	63:08:14.160	&UGS&	3.2 	&	1.8 	&	42.2	&	1.3	&	18.9	&	2.8	&	2.4    	\\
J0333.6+2918	&	03:33:49.004	&	29:16:31.549	&bzb&	5.2 	&	2.1 	&	193.1	&	5.8	&	28.3	&	4.3	&	>4.6	\\
J0334.0+6539	&	03:33:56.738	&	65:36:56.186	&bzb&	8.6 	&	4.2 	&	287.9	&	8.6	&	34.3	&	5.2	&	5.2	    \\
J0338.4+1304	&	03:38:29.279	&	13:02:15.449	&UGS&	10.6	&	5.9 	&	15.1	&	0.6	&	6.13	&	0.94&	>0.5	\\
J0425.3+6320	&	04:25:25.356    &   63:20:01.681	&UGS&	3.9	    &	2.5 	&	24.1	&	1.2	&	0.44	&	0.08&	>0.01	\\
J0515.9+1528	&	05:15:47.353	&	15:27:16.592	&bzb&	6.7  	&	3.6 	&	26.5	&	0.9	&	31.6	&	4.8	&	>5.4	\\
J0516.4+7351	&	05:16:31.232	&	73:51:08.659	&bzb&	4.8	    &	3.4 	&	55.5	&	1.7	&	 8.5	&	1.3	&	>5.5	\\
J0529.0+0937	&	05:29:02.614	&	09:34:31.180	&agu&	7.3 	&	5.0 	&	28.6	&	1.3	&	 0.3	&	0.1	&	3.9	    \\
J0540.5+5822	&	05:40:30.009	&	58:23:38.452	&bzb&	6.5	    &	3.5 	&	29.3	&	1.0	&	11.7	&	1.8	&	>6.6	\\
J0601.0+3838	&	06:01:02.869	&	38:38:29.170	&UGS&	7.3 	&	5.1 	&	704 	&	21  &	62.4	&	9.4	&	92.9	\\
J0605.0+0001	&	06:04:58.418	&	00:00:43.354	&agu&	2.5 	&	1.0 	&	34.4	&	1.1	&	 8.8	&	1.3	&	0.5    	\\
J0605.4+2726	&	06:05:01.141	&	27:24:57.438	&UGS&	4.6	    &	3.2 	&	23.8	&	0.8	&	 2.5	&	0.4	&	>0.2	\\
J0644.2+6036	&	06:44:35.746	&	60:38:51.163	&UGS&	1.9 	&	1.0 	&	33.5	&	1.1	&	 4.1	&	0.6	&	>0.4	\\
J0648.9+1516	&	06:48:47.649	&	15:16:24.800	&bzb&	18.4	&	7.5 	&	64.2	&	2.0	&	37.4	&	5.6	&	7.8	    \\
J0650.4+2056	&	06:50:35.352	&	20:55:57.376	&UGS&	4.5  	&	2.9 	&	 6.4	&	0.5	&	 4.6	&	0.7	&	>3.0	\\
J0706.5+3744	&	07:06:31.697	&	37:44:36.407	&bzb&	10.4	&	4.3 	&	27.0	&	0.9	&	14.7	&	2.2	&	>1.8	\\
J0745.2+7439	&	07:44:05.375	&	74:33:58.255	&bzb&	3.4	    &	1.9 	&	22.6	&	0.8	&	16.3	&	2.4	&	>4.2	\\
J0745.2+8511	&	07:47:16.148	&	85:12:08.614	&agu&	5.1 	&	3.0 	&	11.3	&	0.6	&	 5.0	&	0.8	&	>0.8	\\
J0850.0+4849	&	08:50:00.351	&	48:54:58.671	&bzb&	3.2 	&	1.9 	&	91.6	&	2.8	&	34.6	&	5.2	&	>1.0	\\
J0930.4+8611	&	09:29:43.067	&	86:12:21.292	&bzb&	3.3 	&	1.9 	&	142.3	&	4.3	&	263 	&	40 &	69.3	\\
J0959.5+6535	&	09:58:47.245	&	65:33:54.818	&bzb&	3.1  	&	2.1 	&	729.4	&	21.9&	402 	&	60 &	26.6	\\
J1100.6+4018	&	11:00:21.062	&	40:19:28.050	&bzb&	5.8 	&	2.9 	&	18.0	&	0.7	&	 5.9	&	0.9	&	2.2	    \\
J1107.5+0223	&	11:07:35.916	&	02:22:24.595	&bzb&	2.4  	&	1.3 	&	20.8	&	0.8	&	 7.9	&	1.3	&	>0.3	\\
J1130.4+5814	&	11:31:18.629	&	58:08:58.792	&bzb&	1.5 	&	1.0 	&	42.6	&	1.3	&	21.2	&	3.2	&	>5.0	\\
J1137.0+2553	&	11:36:50.128	&	25:50:52.403	&bzb&	7.2 	&	4.8 	&	14.7	&	0.6	&	 3.3	&	0.5	&	0.1	    \\
J1212.2+2318	&	12:12:38.575	&	23:11:10.826	&UGS&	1.6 	&	0.8 	&	44.0	&	1.4	&	 0.5	&	0.1	&   >0.01	\\
J1223.3+7953	&	12:23:58.075	&	79:53:28.214	&UGS&	2.4 	&	1.4 	&	31.2	&	1.0	&	 8.9	&	1.3	&	>9.4	\\
J1224.5+2437	&	12:24:24.186	&	24:36:23.495	&bzb&	11.5	&	7.0 	&	24.5	&	0.8	&	19.5	&	2.9	&	8.1	    \\
J1244.9+5708	&	12:45:09.999	&	57:09:54.373	&bzb&	1.2 	&	0.7 	&	83.3	&	2.5	&	66.9	&	10.0&	5.5	    \\
J1249.7+3706	&	12:49:46.751	&	37:07:47.913	&bzb&	4.4 	&	2.6 	&	 5.5	&	0.5	&	 5.9	&	0.9	&	1.2	    \\
J1310.9+0036	&	13:11:06.476	&	00:35:10.046	&bzb&	3.9 	&	2.2  	&	17.2	&	0.7	&	15.7	&	2.4	&	3.6	    \\
J1315.0+2346	&	13:14:43.806	&	23:48:26.781	&bzb&	2.0 	&	1.0 	&	183.7	&	5.5	&	60.6	&	9.1	&	>52.8	\\
J1418.6+2539	&	14:17:56.671	&	25:43:26.245	&bzb&	3.2 	&	2.0 	&	88.7	&	3.1	&	10.5	&	1.6	&	>1.8	\\
J1548.3+1455	&	15:48:24.386	&	14:57:02.716	&UGS&	4.2 	&	1.9 	&	23.9	&	0.8	&	17.6	&	2.8	&	>1.0	\\
J1619.8+7540	&	16:19:13.743	&	75:37:53.455	&UGS&	1.4 	&	0.7 	&	88.1	&	2.7	&	15.2	&	2.3	&	0.8	    \\
J1631.0+5224	&	16:30:43.146	&	52:21:38.624	&bzb&	3.1 	&	2.0 	&	120 	&	4	&	18.9	&	2.9	&	>1.6	\\
J1735.9+2033	&	17:36:05.261	&	20:33:01.192	&bzb&	5.9 	&	3.4 	&	24.3	&	0.8	&	 5.8	&	0.9	&	0.3	    \\
J1744.2+1938	&	17:43:57.833	&	19:35:09.017	&bzb&	5.0 	&	3.3  	&	301.2	&	10.6&	105.5	&	15.8&	5.9	    \\
J1809.3+2040	&	18:09:25.452	&	20:41:30.899	&bzb&	3.3 	&	2.2 	&	51.9	&	1.6	&	28.2	&	4.2	&	>4.2	\\
J1841.1+2914	&	18:41:21.720	&	29:09:41.029	&UGS&	4.5 	&	2.5 	&	63.7	&	2.3	&	33.6	&	5.0	&	6.6	    \\
J1841.8+3219	&	18:41:47.042	&	32:18:39.154	&bzb&	4.6 	&	2.1 	&	20.4	&	0.7	&	10.4	&	1.6	&	>4.3	\\
J1926.9+6153	&	19:26:49.891	&	61:54:42.344	&bzb&	10.5	&	2.9 	&	22.1	&	0.8	&	18.5	&	2.8	&	6.8	    \\
J1942.8+1034	&	19:42:47.484	&	10:33:27.101	&agu&	20.7	&	8.1 	&	98.6	&	3.0	&	48.6	&	7.3	&	8.6	    \\
J2002.6+6303	&	20:02:45.347    &   63:02:33.142    &UGS&	2.1	    &	1.3 	&	10.4	&	1.1	&	 1.2    &	0.2&	>0.01	\\
J2004.7+7003	&	20:05:06.013	&	70:04:39.360	&UGS&	6.5 	&	3.2 	&	 6.5	&	0.5	&	 4.8	&	0.7	&	0.1	    \\
J2015.8+3710	&	20:15:28.730	&	37:10:59.515	&bzq&	8.3 	&	3.8 	&	2166	&	65  &	2501	&	250	&  145.5	\\
J2127.8+3614	&	21:27:43.028	&	36:13:05.736	&bzb&	6.1 	&	3.0 	&	191.2	&	5.8	&	38.3	&	5.8	&	3.6	    \\
J2212.5+2803	&	22:12:39.103	&	27:59:38.441	&agu&	1.9	    &	1.0 	&	144.4	&	4.3	&	72  	&	11  &	14.1	\\
J2223.4+0104	&	22:23:29.571	&	01:02:26.650	&bzb&	1.5 	&	1.1 	&	 6.1	&	0.5	&	 6.3	&	1.0	&	1.5    	\\
J2247.7+4412	&	22:47:53.205	&	44:13:15.395	&bzb&	7.8 	&	4.6 	&	70.6	&	2.6	&	28.8	&	4.4	&	2.7	    \\
J2314.0+1446	&	23:13:57.333	&	14:44:23.055	&bzb&	8.8 	&	5.0 	&	40.6	&	1.3	&	14.8	&	2.3	&	0.7	    \\
J2329.1+3754	&	23:29:14.267	&	37:54:14.483	&bzb&	8.4 	&	4.2 	&	19.8	&	0.7	&	13.2	&	2.0	&	3.9	    \\
J2347.2+0707	&	23:46:39.933	&	07:05:06.856	&agu&	6.9 	&	5.0 	&	301.7	&	11.7&	50.5	&	7.6	&	0.6	    \\
\end{longtable}
\end{longtab}
\tablefoot{
\begin{tiny}
\newline
\tablefoottext{a}{For those sources whose radio core is unresolved we calculate lower limits for the brightness temperature.}\\
\end{tiny}
}

\end{document}